\documentclass{article}
\usepackage{arxiv}
\usepackage[utf8]{inputenc} 
\usepackage[T1]{fontenc}    
\usepackage{hyperref}       
\usepackage{url}            
\usepackage{booktabs}       
\usepackage{amsfonts}       
\usepackage{nicefrac}       
\usepackage{bm}
\usepackage{natbib}
\usepackage{lipsum}
\usepackage{graphicx}
\usepackage{amsmath}
\usepackage{graphicx}
\usepackage{amssymb}
\usepackage{algorithm}
\usepackage{algpseudocode}
\usepackage{hyperref}
\usepackage{adjustbox}
\usepackage{amsmath}
\usepackage[usenames,dvipsnames]{xcolor}
\usepackage{graphicx}
\usepackage{amssymb}
\usepackage{booktabs}
\usepackage{algorithm}
\usepackage{algpseudocode}
\usepackage{hyperref}
\usepackage{adjustbox}

\newcommand{\RWnewtext}[1]{{\textcolor{black}{#1}}} %

\def\bSig\mathbf{\Sigma}

\usepackage[english]{babel}
\newcommand{\bch}{\color{BrickRed} }

\newcommand{\ech}{\color{black} \rm}

\newcommand{\mybar}[1]{%
  \mkern 4 mu
  \overline{\mkern-4 mu#1\mkern-0mu}
  \mkern 0mu
}

\newcommand{\LDTFP}{\mbox{LDTFP}}
\newcommand{\DTFP}{\mbox{DTFP}}
\newcommand{\PT}{\mbox{PT}}
\newcommand{\GP}{\mbox{GP}}
\newcommand{\mvPT}{\mbox{mvPT}}
\newcommand{\mvPTGP}{\mbox{mvPT}_{GP}}

\newcommand{\Be}{\mbox{Be}}
\newcommand{\Md}{\mbox{Md}}

\newcommand{\xb}{\bm{x}}
\newcommand{\Zb}{\bm{Z}}
\newcommand{\Cb}{\bm{C}}
\newcommand{\Tb}{\bm{T}}
\newcommand{\Tt}{\tilde T}
\newcommand{\sbf}{\bm{s}}
\newcommand{\veps}{\varepsilon}
\newcommand{\vepsn}{{\varepsilon_d}}
\newcommand{\Bn}[1]{B_{e_1\cdots e_{#1}}}
\newcommand{\BnO}[1]{B_{e_1\cdots e_{#1} 0}}
\newcommand{\BnI}[1]{B_{e_1\cdots e_{#1} 1}}
\newcommand{\en}[1]{e_1\cdots e_{#1}}

\newcommand{\eps}{\epsilon}

\newcommand{\sigs}{\sigma^2}
\renewcommand{\th}{\theta}

\newcommand{\Yepsi}{Y^{(i)}_{\veps}}

\newcommand{\Gbar}{\mybar{G}}
\newcommand{\Mbar}{\mybar{M}}

\newcommand{\GG}{\mathcal{G}}

\renewcommand{\AA}{\mathcal{A}}
\renewcommand{\SS}{\mathcal{S}}
\newcommand{\KK}{\mathcal{K}}
\newcommand{\II}{\mathbb{I}}
\newcommand{\ii}{^{(i)}}
\newcommand{\iip}{^{(i')}}

\title{A Multivariate Polya Tree Model for Meta-Analysis with Event Time Distributions}

\usepackage{setspace}
\author{Giovanni Poli\\
Department of Statistics, Computer Science,
Applications “G. Parenti”, \\
Università degli Studi di
Firenze, Florence, Italy.\\
\texttt{giovanni.poli@unifi.it }
\AND
Elena Fountzilas \\
Department of Medical Oncology, St Luke's Clinic, Thessaloniki, Greece and\\
European University Cyprus, Limassol, Cyprus.\\
\texttt{elenafou@gmail.com}
\AND
Apostolia-Maria Tsimeridou\\
Department of Investigational Cancer Therapeutics, 
The University of Texas MD Anderson Cancer Center, \\
Houston,
TX, USA\\
\texttt{atsimber@mdanderson.org} 
\AND
Peter M\"uller \\
Department of Statistics and Data Science, University of Texas at Austin,  \\
Austin, Texas, U.S.A.\\
\texttt{pmueller@math.utexas.edu}}

\begin{document}

\maketitle

\begin{abstract}
 We develop a non-parametric Bayesian prior for a family of random
  probability measures by extending the Polya tree ($\PT$) prior
  to a joint prior for a set of probability measures
  $G_1,\dots,G_n$,  suitable for meta-analysis with event time
  outcomes. In the application to meta-analysis, $G_i$ is the
  event time distribution specific to study $i$.
  The proposed model  defines   a regression on study-specific
  covariates by introducing increased correlation for any pair of
  studies with similar characteristics.   The desired multivariate
  $\PT$ model is constructed by introducing a hierarchical prior on
  the conditional splitting probabilities in the $\PT$ construction
  for each of the $G_i$. The hierarchical prior replaces the
  independent beta priors for the splitting probability in the PT
  construction with a Gaussian process prior for corresponding (logit)
  splitting probabilities across all studies.
  The Gaussian process is indexed by study-specific covariates, 
  introducing the desired dependence with increased correlation for
  similar studies.
  The main feature of the proposed construction is (conditionally)
  conjugate posterior updating with commonly reported inference
  summaries for event time data.
  The construction is motivated by a meta-analysis over 
  cancer immunotherapy studies.
  \end{abstract}


\section{Introduction}\label{sec:intro}
We introduce a multivariate Polya tree ($\PT$) model for inference
on a set of dependent random distributions  $\{G_i\; i=1,\ldots,n\}$,
suitable for meta-analysis of event time data over multiple
studies -- or cohorts, in the motivating application.
The motivating application is a meta-analysis over
$n$ cohorts in $S$ studies, with each study consisting of multiple patient cohorts
(i.e., $n>S$) and $G_i$ being  the distribution of progression-free survival (PFS) for patients in cohort $i$.
The level of dependence across $G_i$ is modeled as 
a function of
cohort-specific covariate vectors $\xb_i$,
including tumor type, treatment agent, study indicator,  biomarker status
and more. 
We model the dependence between cohort-specific event time distributions $G_i$
by introducing a Gaussian process ($\GP$) prior on the logit conditional
splitting probabilities in the $\PT$ construction. 
We argue that a $\PT$
prior is a natural model for meta-analysis with event-time
outcomes,  which typically report a point estimate $m_i$ for
the median event time and a corresponding confidence interval
$(\ell_i,h_i)$.
We show that the triple $\sbf_i=(\ell_i, m_i, h_i)^\top$ and the sample size $N_i$ (under some
assumptions) are equivalent to reporting counts for the four intervals
defined by $\ell_i, m_i$ and $h_i$. 
An appropriately defined 
$\PT$ prior for such data allows for easy posterior updating, greatly
facilitating inference.
In a very natural and principled way, the proposed model
formalizes the integration of different
sources of knowledge, including data and clinical expert
information. 
Conditioning on the data is implemented through posterior
updating of $G_i$, while 
expert knowledge about the similarity of different cohorts is encoded
in the $\GP$ covariance function. 

Several extensions of the $\PT$ models
to priors for families of random probability measures have been
proposed in the literature.
In the upcoming discussion we will refer to such models generically as
multivariate PT.
Some of the earlier references address the closely related problem of
constructing PT priors for a multivariate distributions, that is,
by way of splitting probabilities for multivariate intervals.
This could in principle be used to define a family of random probability measures
by way of the implied univariate marginals.
For example, \cite{YANG_2008}, use such $\PT$'s 
to define a prior on a bivariate sample space.
\cite{Jara_and_Handson2009} use similar models as a non-parametric
prior for random effects in a semi-parametric regression.
However, this approach is only practicable for a small number
  of random probability measures.
A more general approach is proposed by 
\cite{LiMa2019} who generate dependent random probability measures
by adding an additional 
level in a hierarchical model, with a non-parametric hyper-prior
on the common base measure for multiple $\PT$'s.  
Another approach  proposed in \cite{Trippa_and_Peter}
introduces a gamma process indexed by 
covariates.
Ratios of probabilities under the gamma process define
 marginally beta-distributed splitting
probabilities for dependent $\PT$ priors 
with desired correlation across multiple cohorts
arising from using the same underlying gamma process.  
Specifically in 
the context of meta-analysis, \cite{branscum2008bayesian} developed a
Polya tree mixture model for the random-effects prior. 
\cite{diana2023general} introduce the replicate $\PT$
  framework, which models correlation by imposing constraints on the
  parameters and replicating parts of the trees.
In the approach that we propose in this article, we introduce
correlations using the covariance function of a GP,
directly modeling the correlation
between splitting probabilities that define the cohort-specific event-time distributions $G_i$.
The approach is most similar to the general dependent tail-free process
($\DTFP$) that is defined in \cite{jara2011class}, who then
proceed to propose and implement the special case of the linear dependent tail-free
process ($\LDTFP$).
The $\LDTFP$ uses a normal linear regression for the logit
splitting probabilities.
\RWnewtext{\cite{flores2024clustering} build a model for meta-analysis using
a non-parametric mixture of $\LDTFP$'s}.
Our intermediate model introduced in Section \ref{sec:refPT} is essentially another special case of $\DTFP$ with a tailored prior that allows borrowing information among heterogeneous cohorts.

The proposed construction is motivated by a meta-analysis over $174$
published studies on early-phase cancer immunotherapy.
Immunotherapy has shown promising efficacy results in several types of cancer.
However, only a subgroup of patients benefit from this treatment,
possibly due to patient and tumor heterogeneity.
Depending on the tumor type, approximately
$80\%$ of patients do not respond or even develop hyper-progressive disease (hyper-progression) while a proportion of patients who initially responded eventually develop resistance.
In addition, toxicity remains an issue with some patients developing serious immune-related adverse events. 
Finally, while the use of selected FDA-approved biomarkers is known to be
associated with improved clinical outcomes in selecting patients
receiving immunotherapy \citep{patel2015pd,marabelle2020efficacy},
most immunotherapy trials are still conducted without biomarker
selection. 
These considerations suggest that the use of robust predictive biomarkers
(e.g. gene expression or protein activation) could enable 
 optimal therapy recommendations for patients with diverse
tumor types.  This requires the development of study designs that allow
testing such hypotheses and provide inference on promising biomarkers.
However, published studies are 
systematically underpowered to test hypotheses about biomarker subgroups. 
In this situation, meta-analysis, that is, the pooling of information
across multiple studies, may provide useful.

Standard methods for meta-analysis are based on weighted
  linear regression with random effects  
\citep{Schwarzer2015,Viechtbauer:10,sutton2001bayesian}.
See \cite{ruberu2023meta} for an example of a recent application for
cancer studies reporting relative risks, including a careful
construction to accommodate different reporting modalities across
studies.
They implement meta-analysis using a parametric Bayesian inference model.
Implementations of meta-analysis 
specifically for
survival endpoints are discussed, for example, in
\cite{parmar1998extracting} and \cite{michiels2005meta}. 
\RWnewtext{In particular, \cite{michiels2005meta} discuss meta-analysis
when only the median survival times are reported, including
meta-analysis based on log median ratios across two conditions.
This approach was used in \cite{fountzilas2023correlation}
to analyze same data as in our motivating application.}
We argue that such analysis fails to effectively model the
heterogeneity of the data.  An alternative to account for some of this
heterogeneity could be the use of a parametric meta-regression model,
but the small number of observations (especially for rare tumor types
and less commonly used agents) limits the meaningful use of
meta-regression.  The proposed method introduces a practically
feasible, fully non-parametric alternative in which information
sharing across studies is established in a principled manner within
the framework of an encompassing probability model.

\section{A meta-analysis of cancer immunotherapy studies}\label{sec:data}
We analyze data from a meta-analysis and systematic review
of phase I/II clinical trials assessing the effect of biomarkers on
clinical outcomes
in patients with solid tumors
(Fountzilas et al., 2023a).\nocite{fountzilas2023correlation}
The full data is available from
\nocite{DiB} Fountzilas et al. (2023b).
The analysis did not aim to demonstrate whether specific biomarkers
are predictive of benefit from immunotherapy. 
Such an analysis would be 
of limited validity for rarely evaluated biomarkers.
The goal was to determine whether in general the selection of patients based on biomarkers could be associated with clinical benefit. 
Data were collected using a PubMed search for phase I/II cancer clinical
trials evaluating immune checkpoint inhibitors approved by FDA between
2018 and 2020. 
Only studies that reported summaries
  stratified by biomarker status were selected. 
In total $174$ clinical studies with a total of 19,178
patient responses were included in the analysis in
\cite{fountzilas2023correlation}.
Studies investigated several biomarkers including PD-L1 expression
(111 studies), tumor mutational burden (20 studies) and microsatellite
instability/mismatch repair deficiency (10 studies).

    In this analysis we focus on progression-free survival (PFS) as a
    particular endpoint, which is reported by $S=33$ studies, for a
    total of $n=84$ cohorts.
    { Here, a cohort refers to a subset of patients in a study for which
    results are reported separately, including in particular
    marker-positive and -negative cohorts.
    However, some studies break down results by additional characteristics
    beyond biomarker status, thereby contributing with more than two cohorts.
    The reported summaries for PFS include a point estimate for the median
    PFS ($m_i$) and a corresponding confidence interval
    $(\ell_i,h_i)$ for each cohort, $i=1,\ldots,n$.
    In the proposed inference approach we model the unknown underlying
    distribution $G_i$ of PFS that generated
     event times 
    $y_{ij}$ for $N_i$ patients in cohort $i$. 
    However, posterior updating can only condition on the available summaries
    $(m_i,\ell_i,h_i)$ and the known sample size $N_i$. As in most
    meta-analyses, patient-level data $y_{ij}$ are not available.}

    Let $i^+$ and $i^-$ denote the marker-positive
    and marker-negative cohorts of the same study, that is,
    $\xb_{i^+}$ and $\xb_{i^-}$ differ only by biomarker status.
    Let $G_i^+$ and $G_i^-$ denote the corresponding event time
    distributions with medians $M_i^+$ and $M_i^-$.
    The main inference goal is the comparison of medians, i.e.
    inference about the hypothesis  $M_i^+ > M_i^-$, formalizing the
    motivating question about the use of biomarkers in cancer immunotherapy.
    The proposed multivariate $\PT$ model represents an attractive
    statistical inference approach in this context, as it allows 
    evaluation of the likelihood function for $(m_i,\ell_i,h_i)$ and
    (conditionally) conjugate posterior updating.
    As a side benefit, the borrowing of strength across cohorts
    significantly improves inference for rare conditions, and allows
    more precise inference, for example, for event time distributions
    for patients with rare tumor types, or less commonly used
    treatment agents.

\section{A multivariate Polya tree for event time outcomes}\label{sec:mvPT1}
We first introduce notation by way of reviewing the construction of
a univariate $\PT$ prior in Section \ref{sec:PT1}.
In Section \ref{sec:refPT}  we
review the construction of the $\DTFP$ of \cite{jara2011class} which
extends the construction to a multivariate $\PT$ 
with common partitioning subsets, which is then finally, in Section
\ref{sec:mvPT2}, extended to allow for 
different partitioning subsets for each distribution. 
The latter is needed for the desired meta-analysis with event
time data. 

\subsection{Univariate Polya Trees}
\label{sec:PT1}

The $\PT$ \citep{lavine1992some,lavine1994more} is a prior distribution for a
random probability  measure $G$ defined over a sample space $S$. 
It is
constructed recursively using nested partitions
$\pi_d = \{\Bn{d};\; e_\ell \in \{0,1\}, \ell=1,\ldots,d\}$,
$d=1,2, \ldots$,
of the sample space, starting with $S=B_0 \cup B_1$ and
recursively refining the partition
with $\Bn{d} = \BnO{d} \cup \BnI{d}$.
That is, $\BnO{d}$ and $\BnI{d}$ are defined by splitting
$\Bn{d}$ into a left and right binary partitioning subset.
Following standard notation, we use subscript $0$ for left and $1$ for
right partitions.
We use $\veps_d=\en{d}\ \in \{0,1\}^{d}$ to uniquely identify a partitioning
subset $B_\vepsn \in \pi_d$.
A prior model on $G$ is implicitly  defined by 
a prior on the conditional splitting probabilities
$Y_{\veps_d0} \equiv G(B_{\veps_d0} \mid B_{\veps_d})$,
together with the choice of the partitioning subsets $B_{\veps_d}$.
The standard $\PT$ prior assumes 
$Y_{\veps_d0} \sim \Be(\alpha_{\veps_d0}, \alpha_{\veps_d1})$
(and $Y_{\veps_d1}=1-Y_{\veps_d0}$).
The construction can be described as a
sequence of increasingly refined random histograms,
with bins defined by $B_{\veps_d}$ and corresponding probabilities
$G(B_{\veps_d}) = \prod_{\ell=1}^d Y_{e_1\cdots e_\ell}$, and is illustrated in Web Figure 5 in the on-line supplementary materials.

The construction defines a random
probability measure $G \sim \PT(\AA, \Pi)$, where $\AA=\{\alpha_{\varepsilon_d0},\alpha_{\varepsilon_d1}:\  d=1,2,\ldots\}$
is the set of hyper-parameters that index the beta priors on $Y_{\veps0}$ and
$\Pi = \{\pi_d :\ d=1,2,\ldots\}$ is the nested partition sequence.
The hyperparameters $\AA$ and $\Pi$ can be chosen to ensure a
desired prior mean, $\mathbb{E}[G]=G_0$. 
Expressing prior information by way of a centering distribution or
prior mean  is a common feature in nonparametric Bayesian models. 
In the PT model there are two main strategies
to achieve a desired prior centering.  The first is to fix $\Pi$
and adjust the hyperparameters
$(\alpha_{\varepsilon_d0},\alpha_{\varepsilon_d1})$ to ensure
$\mathbb{E}[Y_{\veps_d0}] = G_0(B_{\varepsilon_d0}\mid
B_{\varepsilon_d})$.
The alternative strategy is to fix $\AA$ to ensure
$\mathbb{E}[Y_{\varepsilon_d0}]=0.5$ and then achieve the desired prior
centering by using dyadic quantiles under $G_0$
as partitioning subsets, i.e.;
  defining $B_{\veps_d0}$ such that $G_0(B_{\veps_d0} \mid B_{\veps_d}) = 0.5$
\citep{lavine1992some, lavine1994more}.
A common choice for $\alpha_\vepsn$ in the latter case is 
$\alpha_{\veps_d0}=\alpha_{\veps_d1} = c\cdot(d+1)^2$, which ensures a continuous
random distribution under a PT prior.
The hyperparameter $c$ is a scalar precision  parameter,
which is widely discussed in the $\PT$ literature.
We will later use both strategies. We use fixed partitioning subsets for levels $d=1,2$ (matching the intervals defined by $\ell_i, m_i$ and $h_i$),  and dyadic splits for $d>2$.
$\PT$ priors have several attractive and useful properties, including
conjugacy under i.i.d. sampling and flexibility in
encoding prior beliefs.
Compared to other non-parametric priors, the main drawback of the PT
model is the lack of smoothness of the density of $G$. However, this
is of less concern in survival analysis, as the primary target is
often the cumulative density function or, equivalently, the survival
function shown in the Kaplan-Meier curve.
  
\subsection{Multivariate Polya tree with
  Gaussian process dependence} 
\label{sec:refPT}

\cite{jara2011class} introduce a prior for a family of random
probability measures   $\{G_i;\; i=1,\ldots,I\}$ based on a
generalization of a univariate $\PT$ prior.
We review their construction, introducing some variations in
anticipation of the next extension.
In all variations, $G_i$ remains a random probability measure over
a sample space $S$ with (at least approximately) a marginal $\PT$ prior.

Recall that in the motivating application $n$ is the number of cohorts
with available data  on PFS.
In anticipation of posterior predictive inference for future studies,
we are setting up the model for $I>n$ cohorts, including cohorts indexed
by $i\in\{n+1,\dots, I\}$ without observed data. 
We first set up a model 
sharing a {\it common partitioning sequence} $\Pi$ across all $i$. 
Let then $ Y\ii_{\veps_d0}= G_i(B_{\veps_d0} \mid B_{\veps_d})$  denote the splitting
probabilities under $G_i$ and let
$\eta(p)=\log\{\ p\ (1-p)^{-1}\}$ denote a logistic link function.  
The model maintains independence of splitting probabilities $\Yepsi$
for different partitioning subsets $B_\veps$ within the same tree, but
allows the splitting probabilities for the same $\veps$ to be
correlated across $i$. 
This is achieved by introducing a Gaussian process (GP) prior on 
$Z_{\varepsilon0}\ii= \eta(Y_{\varepsilon_d0}\ii)$. 
The GP is indexed by cohort-specific covariates $\xb_i$ and
replaces the independent beta prior of the univariate PT
construction. 
That is, we assume $\{Z\ii_{\veps_d0}\}_{\xb_i \in X} \sim
\GP(\mu_{\veps_d0},K_{\veps_d0})$  with mean function $\mu_{\veps0}$ and covariance
function $K_{\veps_d0}$ (see below for $\mu_{\veps_d0}$ and $K_{\veps_d0}$).
There is a separate, independent GP prior for each $\veps_d0
=\eps_1\cdots\eps_d0$
 (and recall that any $Y\ii_{\veps_d1}$ is implied as
  $1-Y\ii_{\veps_d0}$).
Dependence is limited to partitioning subsets up to a certain
depth $D$ of the tree,  with GP priors for each
$\veps_d0$ up to level $D$, and independent beta priors beyond.
We write $(G_1,\dots,G_I) \sim \mvPT_{GP}(\Pi,D,\KK,\AA)$.
The parameter $\KK$ is the set of $2^D-1$ pairs of mean and covariance
functions
$\big(\mu_{\veps_d0}(\cdot);K_{\veps_d0}(\cdot,\cdot)\big)$
that define the $\GP$ priors.
The parameter $\Pi$ is the (common) nested partition sequence.
Finally, the set $\AA$ is defined as in the univariate $\PT$ prior
and collects the parameters $\alpha_{\veps_d0},\alpha_{\veps_d1}$ for 
$d > D$.
Posterior updating is similar to the  univariate  $\PT$ and
depends on the counts of observations in each sub-interval of
$\Pi$.  
Logit splitting probabilities $Z\ii_{\veps_d0}$ associated with the same
sequence $\veps_d0$ are dependent across $i$ (i.e., cohorts) and it is convenient to
sample them jointly. 
Using a logit link allows easy posterior updates using the Polya-Gamma
sampler introduced by \cite{polson2013bayesian}.
The same tree-based logit normal is introduced
  as logistic-tree normal in \cite{wang2022microbiome} and
  \cite{LDAlogitlima} where it is used as a prior 
  for categorical probabilities in a mixed membership model.

For the hyperparameters we propose choices that imply
  a desired marginal distribution for $G_i$, similar to the univariate PT.
  For details see below (in the context of setting up marginal moments
  $\mu_{\veps_d0}$ and $\sigs_{\veps_d0}$ for the GP prior).
For the shared nested partition sequence $\Pi$ we
use the dyadic quantiles of a base measure $G_0$.
Additionally, for $\veps_d \in \{0,1\}^d$ at 
levels $d=1,\ldots,D$ we need a covariance function
$K_{\veps_d0}(\cdot,\cdot)$ and mean process $\mu_{\veps_d0}(\cdot)$ for
the $\GP$ prior. We factor the covariance function as
$K_{\veps_d0}(\xb,\xb') = \sigs_{\veps_d0} \cdot R(\xb,\xb')$, and
first consider the moments of the marginal distribution
$Z\ii_{\veps_d0} \sim N(\mu_{\veps_d0},\sigs_{\veps_d0})$.
Our choice is based on fixing the parameters
$(\mu_{\veps_d0},\sigs_{\veps_d0})$ to approximately match a
$\Be(\alpha_{\veps_d0}, \alpha_{\veps_d1})$ prior on
$\eta^{-1}(Z\ii_{\veps_d0})$, that is, the prior under a univariate
$\PT$. 
Although beta and logistic-normal distributions are never
exactly equal, any beta distribution can be approximated with the 
logistic-normal distribution that minimizes Kullback-Leibler
divergence \citep{logitnormal_10.2307/2335470}.  
Let $\psi$ and $\psi'$ denote the digamma and trigamma functions, respectively.
We use $\mu_{\veps_d0}(\xb)=\psi(c\cdot [d+1]^2)-\psi(c\cdot [d+1]^2)=0$
and $\sigma^2_{\veps_d0} =2\cdot\psi'(c\cdot [d+1]^2)$.
%
Having specified the marginal moments $\mu_{\veps_d0}$ and
$\sigs_{\veps_d0}$,
we are left to specify $R(\xb,\xb')$,
i.e., the correlation between logit-transformed
splitting probabilities 
for cohorts with covariates $\xb$ and $\xb'$.
The correlation function $R(\xb,\xb')$  is used
to introduce clinical expert judgment on similarity of the event time
distributions $G_i$. See Section \ref{sec:cov} for an example of
constructing $R(\xb,\xb')$ tailored to our application. 
For levels $d> D$ we define partitioning subsets $B_{\veps_d0}$ and
beta parameters $\alpha_{\veps_d0}, \alpha_{\veps_d1}$ to achieve a
desired prior mean $G_0$ as described in Section \ref{sec:PT1},
using $\alpha_{\veps_d}=c
 d^2$.
In summary, 
$R(\xb,\xb')$ introduces clinical expert judgment on how similar
event time distributions for different cohorts
are likely to be, and the precision parameter $c$ and the centering
measure $G_0$ fix prior uncertainty and expectation of the marginal
prior on $G_i$.
Defining prior elicitation for the GP parameters by approximating
the beta prior in the standard PT construction
allows us to use the same $c$ and $G_0$ to characterize splitting
probabilities across all levels, including $d=1,2$ with the
logit-normal GP prior as well as $d>2$ with the beta prior.
 
\begin{figure}
\centering
  \includegraphics[width = .85\textwidth]{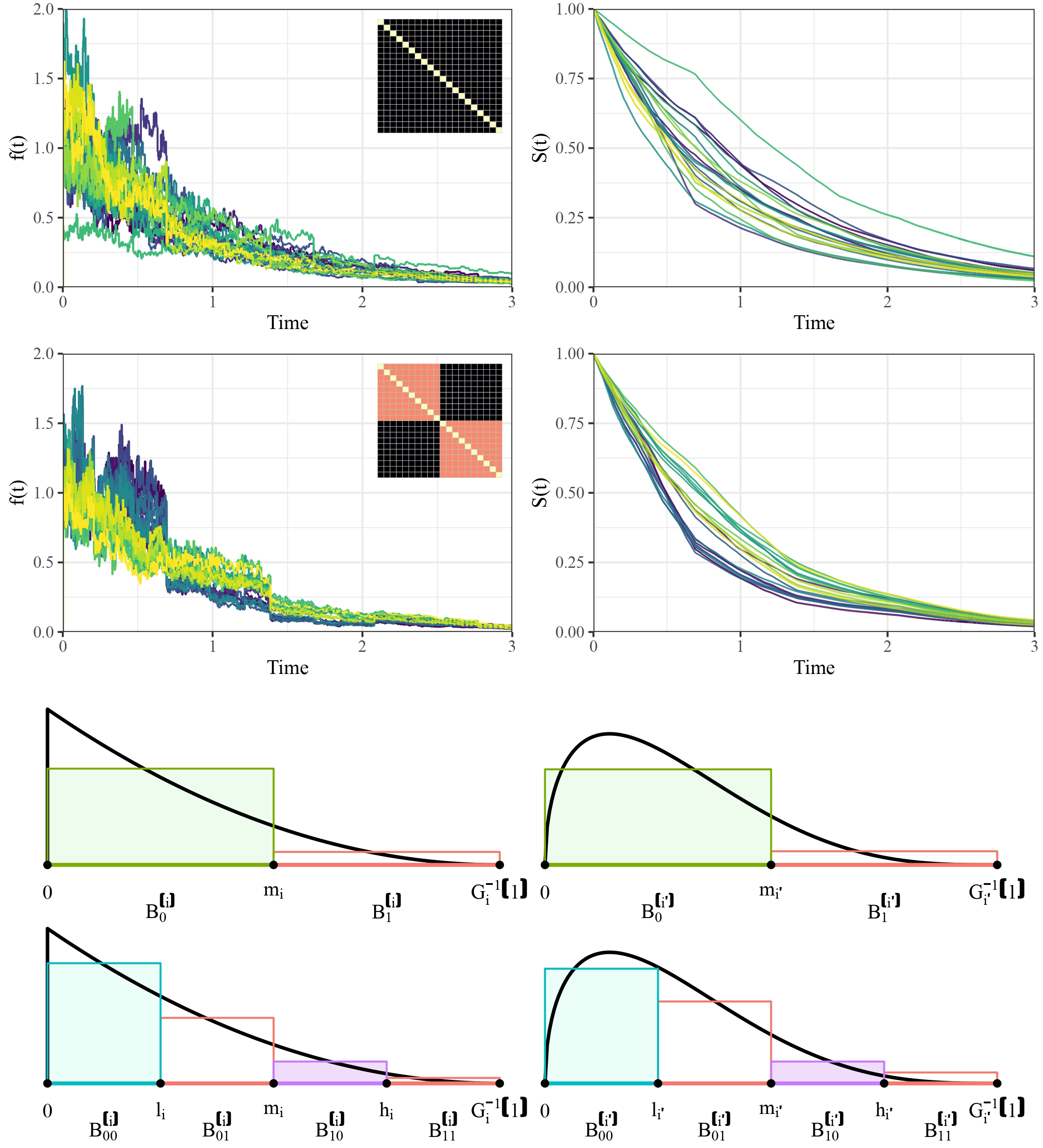} 
    \caption{The top four panels show random samples $(G_1,\ldots,G_n)$  from a
      $\mvPT$ prior, using two different choices of $R(\xb,\xb')$
      (plotted in the square insert in the left panels).
      The left panels show the densities.
      The right panels show the survival functions. Both use
      $G_0=\mbox{Exp}(x\mid\lambda=1)$ and precision parameter $c=5$.
      For the first \bch row, \ech $R(\xb,\xb')$
      is the identity (implying in particular that $G_i$ are
      exchangeable), in the second \bch row \ech it
      is block diagonal (the $G_i$ are partially exchangeable within
      each block - e.g., marker-positive and -negative cohorts).
      The bottom panels show the random probabilities
      $G_i(B\ii_{\varepsilon_d})$ for $d=1,2$ (in the bottom two rows) for
      $i \ne i'$ (in the two columns).
      Splitting probabilities $Y\ii_{\veps_d}$ for subsets $B\ii_{\veps_d}$ marked with the same color (except red) are dependent (across $i$), red-colored bins are deterministic given the other bins (i.e., $Y_{\varepsilon_{d}0}\ii$ implies $Y_{\varepsilon_{d}1}\ii$).
      The overlaid density curve shows the density
      defined by the limit as $d \to \infty$.}
    \label{fig:sim}
\end{figure}

If desired it is possible to define hyperpriors and
potentially learn about parameters in the covariance functions
\citep[Chapter 15]{MurphyGP}.
However, this possibility is not explored here.
Finally, let $B \subset A$ denote any two nested
subsets, and let $\boldsymbol{Z}=\{Z\ii\}_{i=1}^I$
with $Z\ii = \eta\{ G_i(B \mid A)\}$.
Via Monte Carlo prior simulation the proposed model allows to evaluate
$\mathbb{E}[Z\ii]$ and $\mathbb{C}ov(Z\ii,Z\iip)$ for any $i,i'$. 
We shall use this later.
Pseudo code for this prior simulation is available online as supplementary materials.
Figure \ref{fig:sim} shows a random sample $(G_i, i=1,\dots,20)$ from
a $\mvPT$ for two different correlation matrices $R(\xb,\xb')$, and
illustrates how two marginal random distributions are constructed from
$G_i(B\ii_{\veps_d0} \mid B\ii_{\veps_d})$.

\section{A Polya tree prior for meta-analysis with event time data}\label{sec:mvPT2}

\subsection{
  Multivariate Polya tree with study-specific partitions}
\label{subsec:ExmvPT} 

Recall the format of the data with $\sbf_i=(\ell_i,m_i,h_i)$ and
 sample size $N_i$ for each cohort in the meta analysis. 
We assume that $m_i$ and $(\ell_i,h_i)$ were
 determined as the intersections of the
Kaplan-Meier ($\mbox{KM}$) survival curve \citep{kaplan1958nonparametric}
and the corresponding error bounds, respectively, with the 
0.5 threshold.
We asume \ that the error bounds are based on the $\mbox{KM}$
estimator and the Greenwood formula
\citep{greenwood1926natural,hosmer2011applied}.
Conditioning on a censoring pattern (i.e., the order in which observed
event times and censoring events occur -- see below about updating
this assumption) $\sbf_i$ implies counts of observations in each of
the four sub-intervals determined by $(\ell_i, m_i, h_i)$.  If these
subintervals match the partitioning subsets in the first two levels of
the marginal $\PT$ construction, then the counts are a sufficient
statistic for the posterior distribution of $Y_0\ii, Y_{00}\ii,
Y_{10}\ii$, $i=1, \ldots, n$, i.e., to update knowledge on $G_i$.
We  therefore  replace the shared partition sequence $\Pi$ of the
$\mvPT_{GP}(\Pi,D,\KK,\AA)$
by a set of {\it cohort-specific partition sequences} $\{\Pi_i\}_{i=1}^{I}$, with
$\pi\ii_1=\{[0,m_i),[m_i,+\infty)\}$ and
$\pi\ii_2=\{[0,\ell_i),[\ell_i,m_i), [m_i,h_i), [h_i,+\infty)\}$.
Nested partitions $\pi_d$  at deeper levels $d>2$  are
constructed by dyadic splits of the parent set $B\ii_{\veps_{d-1}}$ such that $G_0(B\ii_{\veps_d}\mid B\ii_{\veps_{d-1}})=0.5$. 
We refer to the extended model as
$\mvPTGP(\{\Pi_i\}_{i=1}^I, D,\KK,\AA)$, with $\{\Pi_i\}_{i=1}^I$ replacing
the common shared partitioning sequence $\Pi$ of the earlier
construction.
The extension requires careful consideration
of the mean process $\mu\ii_{\veps_d0}$ and the covariance function
$K\ii_{\veps_d0}$. Note the added superindex for cohort $i$, to
allow for different $B\ii_{\veps_d}$.
The elicitation involves expected values
 and covariances for each (logit transformed)
conditional probability $G_i(B\ii_{\veps_d0} \mid B\ii_{\veps_d})$ and 
$G_{i'}(B\iip_{\veps_d0} \mid B\iip_{\veps_d})$,
which now refer to possibly very different sets $B\ii_{\veps_d0}$
and $B\ii_{\veps_d0}$.
A principled and coherent specification of such quantities
is challenging.
We use the following construction to reduce the problem to the earlier
case of shared $\Pi$. 
We first consider a process with shared partitions   $\GG^* = (G^*_i,
i=1,\ldots,I) \sim \mvPTGP(\Pi, D, \KK, \AA)$,
defined as in the previous section.
Under $\GG^*$ we can then by the earlier discussed prior simulation
evaluate probabilities for any events.
In particular, we can evaluate expected values and covariances for
logit conditional probabilities $G_i(B\ii_{\veps_d0} \mid B\ii_{\veps_d})$,
as needed for the construction of $\mvPTGP(\{\Pi_i\}_{i=1}^I,D,\KK,\AA)$.
  
Finally, we have to select partition sequences $\Pi_i$ for future cohorts ($i>n$).
We proceed with $\ell_i$ defined as the median of $\ell_k$,
$k=1,\dots,n$, and similarly for $m_i$ and $h_i$.

\subsection{ Censoring patterns and posterior inference}
One feature of the proposed model is that it allows to condition on
all available information, beyond only the median point estimates.
Instead, we condition on  the
entire reported triple $\sbf_i$, including confidence intervals (if
available). 
To map $\sbf_i$ to counts used for posterior updating of $\PT$
parameters, we need to make assumptions about the censoring pattern (i.e., the sequence
of observed and censored event times).
We start by assuming a distribution for censoring times, $C\ii_j
\sim H$  for patient $j$ in study (cohort) $i$. 
To update an assumed censoring pattern we employ an ABC-like
\citep{marin2012approximate} Metropolis-Hastings scheme.
Specifically,
we start with an initial assumption for the censoring pattern and then update following  Metropolis-Hastings transition probabilities, accepting only transitions
that accurately reproduce the observed triple $(\ell_i,m_i,h_i)$. The latter is assumed to be derived from a Kaplan-Meier estimator.
We refer to the simulation as an ABC algorithm since we generate new values for $C\ii_j$ and $T_j\ii$ 
accepting only those that imply a match with the reported statistics
$\boldsymbol{s_i}$.
See Appendix B for details of this simulation, including the
  initialization. 

Keep in mind that this simulation is only imputing censoring patterns
- there is no notion of posterior simulation of parameters.
This simulation-based approach enables us 
to accommodate different levels of information provided for each
study.  
For example, for a study that reports no censoring events we use 
deterministic counts, while a study that reports the number of censoring
events can be treated differently from a study that reports no
details on censoring. 
Finally, for a study that reports confidence intervals for different
coverage probability than others, it is straightforward 
to account for this choice  in the derivation of point estimate
and confidence intervals for the median from the Kaplan-Meier plot.
  

\subsection{Posterior summaries}
\label{subsec:inf}
 

Recall that $G_i$ is the distribution of event times in cohort $i$,
with covariates $\xb_i$ and that the model is jointly defined on
$G_1,\dots, G_n$ ($i=1,\ldots,n$ are the cohorts with observed data) and $G_{n+1},\dots,
G_I$ (future cohorts).  
Let $\Md(P)$ be the median of a probability measure P, let $M_i=\Md (G_i)$
denote the median of $G_i$ and let
$\SS=\{\boldsymbol{s_i}\}_{i=1}^n$ denote the observed data.
We suggest reporting $p(M_i\mid\SS)$ to summarize inference under the proposed model. 
If point estimates are needed, we use the  posterior median  $\widehat{M}_i = \Md\big\{p(M_i | S)\big\}$. 
For the observed cohorts, $i=1,\ldots,n$, the posterior distribution
$p(M_i\mid\SS)$ 
summarizes updated knowledge.
However, from an inferential point of view, the main interest
is on the posterior distribution $p(M_i\mid\SS)$ for future cohorts $i=n+1,\dots,I$. 
Moreover,  research often focuses on populations that correspond
to multiple covariate vectors $\xb_i$, i.e.; mixtures of multiple
future cohorts $i$ with $i\in A \subseteq \{n+1,\dots, I\}$.
For example, inference on marker-positive patients is naturally
represented as a mixture where $A$ is the set of all marker-positive
future cohorts. 
Event time distributions for such populations are implicitly defined
as $P=\sum\limits_{i\in A} \pi_i\ G_i$, weighting different cohorts
with possibly non-uniform weights $\pi_i$.

One aspect to keep in mind with inference on $M_i$, $i>n$, is
that $p(M_i \mid \SS)$ also includes study-to-study  variation. 
In contrast, inference under classical meta-regression usually
reports p-values for fixed effects $\th$, that is, an average effect
for future studies with particular characteristics $\xb$. 
To define a comparable inference summary under the proposed
non-parametric model,
let $\th=\{Y^{(h)}_{\veps_d}: d=1,\ldots\ ; \ h=1,\ldots,n\}$
denote the conditional splitting probabilities for the observed
studies and define 
$\Gbar_i = \mathbb{E}[G_i \mid \th]$ and $\Mbar_i = \Md(\Gbar_i)$, $i=n+1, \ldots, I$.
Then $\Mbar_i$ is a function of $\theta$ and $p(\Mbar_i \mid \SS)$
reports uncertainty on the future cohorts without cohort-to-cohort variation. 
In the upcoming results, we report summaries of $p(\Mbar_i \mid \SS)$ or
credible intervals based on it as alternatives to 
$p(M_i \mid \SS)$. 

Recall the definition of $G_i^+$, $G_i^-$, $M^+_i$ and $M^-_i$ as the
event time distributions and corresponding medians
for matching marker-positive and marker-negative cohorts.
Similarly, let $P^+$ and $P^-$ and corresponding medians $M_P^+$, $M_P^-$ refer to mixture populations 
 differing only by the presence of biomarkers. 
Inference  summaries reported in the upcoming discussion
will focus on the paired comparison of $M^+_i$ vs. $M^-_i$ and $M_P^+$
vs. $M_P^-$. 
Alternatively we will also report similar quantities for the expected
values, ${\Mbar_i}^+$ vs. ${\Mbar_{i}}^-$ and $\Mbar_P^+$
vs. $\Mbar_P^-$.

\section{Correlation function and prior specification}\label{sec:cov}
We describe the elicitation of the $\mvPTGP$ prior that is
used for the motivating meta-analysis problem and in the
  simulations, with minor adjustments. 
In the studies under consideration median progression-free survival
(PFS) is usually reported for within the first few months.
We therefore use a half-Cauchy with $\sigma = 3.5$ (in months)
as centering distribution $G_0$ to allow a heavy right tail.
Next, we fix $c$ at a  weakly informative value
$c=5$, which implies an \textit{a priori} 95\% credible interval of
$[1.08; 11.52]$ on the median for a new study $G_i$.

We construct the correlation function $R(\xb,\xb')$ to represent
clinical judgment about the level of similarity between
the event time distributions for any pair of cohorts $i, i'$.
We first describe the construction of $R(\xb,\xb')$ for any two
cohorts with matching biomarker status.
We proceed by introducing an additive similarity score, adding points
for each matching categorical covariate $x_j=x'_j$, which is then rescaled to
a unit maximum. The covariates  that are expected to have
  the strongest association with PFS are tumor type and agent.
If cohorts $i$ and $i'$ share either of these two characteristics
we record an increment of 2 points each for the similarity score.
Tumor type includes \textit{`other'}. 
Two cohorts with \textit{`other'} as the tumor type are considered unmatched.
Covariates  that are  judged to be less likely  strongly
  associated with PFS 
include biomarkers status, study phase (1, 1/2, or 2), line of therapy
($1$, $\ge2$, or any) and type of therapy (combination,
combination-or-monotherapy or monotherapy).
For each matching secondary covariate, we record a 0.5 increment of
the similarity score. 
Next, 1 point was added for cohorts $i$, and $i'$ that are part of the same study,
and an extra point is added on the diagonal for $i=i'$ to add a nugget
in the implied covariance function.
For cohorts $i, i'$ with different biomarker status we only
  apply the rule about the shared study, recording a similarity score
  of 1 point for cohorts within the same study.
The reduced correlation for cohorts with different biomarker status
avoids over-smoothing of biomarker effects.

The described construction implies a maximum 
similarity score of 8. Rescaling to a maximum of 1 defines
$R(\xb,\xb')$. 
In summary,  letting  $\omega_j$  denote the previously introduced
covariate-specific weights, and
assuming that   $x_{i1}$ is a study indicator, and $x_{i2}$ is an
indicator for biomarker status,
we define $R(\xb_i,\xb_{i'})$ for
$\xb_i=(x_{i1},\dots,x_{iJ})$ and
$\xb_{i'}=(x_{i'1},\dots,x_{i'J})$ as follows: 
\begin{equation}
  R(\xb_i,\xb_{i'})= \frac{
    \mathbb{I}(i=i') +
    \omega_1 \cdot\ \mathbb{I}(x_{i1}=x_{i'1}) +
    \mathbb{I}(x_{i2}=x_{i'2}) \cdot  \sum_{j=2}^J \omega_j\cdot\ \mathbb{I}(x_{ij}=x_{i'j}) }
   {1+\sum_j \omega_j}.
    \label{cor_fun}
\end{equation}
See Appendix C for an argument that $R$ defines a positive
semi-definite correlation matrix, and for suggested adjustments if
a similar construction does not define a positive definite matrix.
In our application we apply \eqref{cor_fun}  only for categorical covariates, but
  a similar kernel construction can be adapted for continuous variables.
  Also, keep in mind that $R(\xb,\xb')$ is only used to construct a
  valid $I \times I$ covariance matrix for the given studies. The
  emphasis is on constructing a suitable $R(\xb,\xb')$ to represent
  prior judgment. 

The level of censoring for PFS is only moderate.
We assume an exponential distribution with mean 10 for the censoring
distribution in all cohorts.

\section{Simulation study}\label{sec:sim}
The simulations aim to assess inference in a realistic
  scenario which calls for standard meta-analysis, and to show that the assumed
  dependence across cohorts can compensate for the more restrictive
  borrowing of strength under a parametric model. 
We construct a simulation truth to mimic the setup in the motivating
study, using the Kaplan-Meier estimator to generate simulated datasets
of summaries $\sbf_i=(\ell_i, m_i, h_i)$ and for each hypothetical
dataset, we compare inference under the $\mvPT$ model versus standard
methods for meta-analysis.

We assume three tumor types ($TT1, TT2, TT3$), and
two agents $(A0, A1)$, with $A1$ being associated with higher median PFS.
The resulting six cases were further divided on the basis of biomarker
status (positive or negative),  resulting in 12 types of cohorts
shown in Table \ref{tab.sim}. Biomarker-positive status is 
assumed to add a positive offset of 0.5 to the median. 
Simulating a certain number of studies (see Table \ref{tab.sim})
for each of the combinations of tumor and agent
we  generated a total of 25 hypothetical studies,
with a marker-positive and a marker-negative cohort for each study,
resulting in a total of $n=50$ cohorts for each simulated dataset.
Finally, for each study we generate a
study-specific multiplicative random effect on the median,
$\alpha_i\sim\mathcal{U}(0.8,1.2)$.  
See Table \ref{tab.sim} for the assumed true medians (before applying
the study-specific random effects) for each of the $12$ unique
combinations of covariates.
We then simulated event times for $N_i=20$ subjects for each of the $50$ cohorts,
using the distributions indicated in Table \ref{tab.sim}.
Finally, using the \texttt{survfit} function in
the {\em R} package {\em survival} \citep{survival-package}, we
evaluated point estimates $m_i$ and corresponding 95\% confidence
intervals $(\ell_i,h_i)$ for median PFS for each
cohort. the triples $\sbf_i$ are the data.
For simplicity, we omitted censoring.

\begin{figure}
    \centering
    \includegraphics[width = \textwidth]{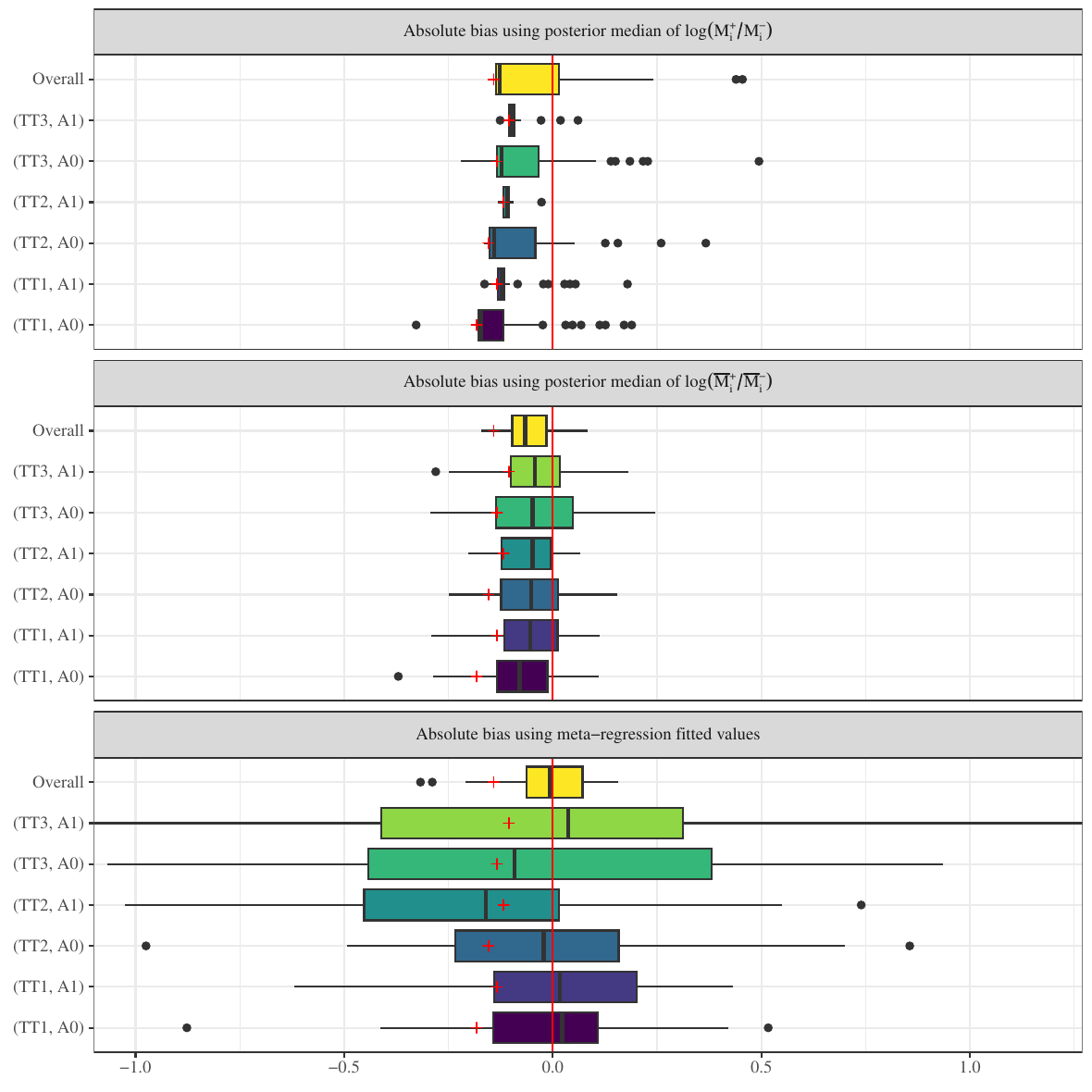}
    \caption{Box-plots of realized bias over 50 repeat simulations, for
      different point estimates.
      The top panel shows bias for the posterior median of $\log\left(M^+_i/M^-_i\right)$.
      The middle panel shows bias  for the posterior median of
      $\log( {{\Mbar_i}^+} /{{\Mbar_i}^-})$. 
      The bottom panel shows bias for 
       log median ratios under meta-regression approach. 
      For meta-regression bias are calculated using the fitted expected values.
      In all three plots the red $+$ marks the bias of prior
      expected values (i.e., $\log(1)=0$). The shifts are due to the different simulation truths of the different covariate combinations.}
    \label{fig:BIAS}
\end{figure}

We used the same prior distributions as for the data analysis with
the real data, namely 
$c = 5$ and a half Cauchy with scale $3.5$ for $G_0$.
We construct a correlation function $R(\xb,\xb')$ as in Section
\ref{sec:cov}, using only the rules for tumor type, agent and
biomarker status.
We add future cohorts, $i=n+1,.., I$, including one cohort for each
of the 12 unique combinations of covariates, i.e., $I=n+12$.
The choice of $R(\xb,\xb')$ 
is an important step in the model construction,  representing
informative prior expert judgment. To explore the impact of this
choice we carried out some sensitivity analysis with alternative
correlation functions on the same simulated data.
These simulations are summarized in the online supplementary materials.
The simulations show noticable differences in
inference under alternative constructions, confirming the importance
of an expert-informed construction, as described.
To summarize posterior inference on specific tumor-agent pairs,
we evaluate log ratios of median PFS, $\log (M_i^+/M_i^-)$, for the 12
future cohorts, comparing marker-positive and -negative pairs
$(i^+,i^-)$ of cohorts with matching tumor type and agent.
We also evaluate an overall effect of biomarker status by
considering a mixture of the 6 marker-positive ($P^+$) and the
6 marker-negative ($P^-$) future cohorts using weights
proportional to $S_i$ in Table \ref{tab.sim}.
We evaluate point estimates as posterior medians of
$\log (M_i^+/M_i^-)$ and $\log (M_P^+/M_P^-)$, respectively.
\RWnewtext{We used  log median  ratios because
  those are used in the classical meta-regressions that we report
  for comparison.
  For the latter we used a random effects model with just the intercept (i.e., that estimates the overall effect) and a meta-regression
  including an interaction of tumor type and agent (i.e., including a parameter for each the six combination). }
The data are the log ratios of median survival times reported
  in the studies \citep{michiels2005meta}.
\RWnewtext{For both models} we 
  used the implementation in the \texttt{R} package 
\texttt{metafor} \citep{Viechtbauer:10}.
\begin{table}
  \caption{Simulation setup. The table shows the assumed cohort-specific
    medians. Each event time distributions are parameterized in terms of the
    median $\lambda$. Parameters $\alpha_i\sim\mathcal{U}(0.8,1.2)$ are
    study-specific multiplicative effect on the median.
    $S_i$ reports the number of simulzated studies sampled with each distribution.}
  \label{tab.sim} 
\resizebox{\textwidth}{!}{\begin{tabular}{c c c c}
\toprule
\multicolumn{1}{c}{ }& \textbf{Tumor Type 1} & \textbf{Tumor Type 2} &  \textbf{Tumor Type 3}\\
\multicolumn{1}{c}{ }&  $\mathbf{Exp}(\lambda)$ & $\mathcal{HN}(\lambda)$ &  $0.5\ \mathbf{Exp}(\lambda)+0.5\ \mathcal{HN}(\lambda)$\\[0.5ex]
  \cmidrule(lr){2-2}\cmidrule(lr){3-3}\cmidrule(lr){4-4}
\textbf{A0} &
  $ \begin{aligned}
  \lambda_i^+ &= (2.5+0.5)\cdot\alpha_i\\[0.5ex]
  \lambda_i^- &= 2.5\cdot\alpha_i\\
   & & S_i=5\\
  \end{aligned} $
  & 
  $ \begin{aligned}
  \lambda_i^+ &= (3+0.5)\cdot\alpha_i\\[0.5ex]
  \lambda_i^- &= 3\cdot\alpha_i\\
     & & S_i=5\\
  \end{aligned} $ 
  &
  $ \begin{aligned}
  \lambda_i^+ &= (3.5+0.5)\cdot\alpha_i\\[0.5ex]
  \lambda_i^- &= 3.5\cdot\alpha_i\\
  & & S_i=3\\
  \end{aligned} $ \\
 \cmidrule(lr){2-2}\cmidrule(lr){3-3}\cmidrule(lr){4-4}
\textbf{A1}&
  $ \begin{aligned}
  \lambda_i^+&= (2.5+0.5+1)\cdot\alpha_i\\[0.5ex]
  \lambda_i^-&= (2.5+1)\cdot\alpha_i\\
     & & S_i=5\\
  \end{aligned} $ 
& \multicolumn{1}{c}{
  $ \begin{aligned}
  \lambda_i^+&= (3+0.5+1)\cdot\alpha_i\\[0.5ex]
  \lambda_i^-&= (3+1)\cdot\alpha_i\\
     & & S_i=5\\
  \end{aligned} $}  &
  \multicolumn{1}{c}{
  $ \begin{aligned}
  \lambda_i^+&= (3.5+0.5+1)\cdot\alpha_i\\[0.5ex]
  \lambda_i^-&= (3.5+1)\cdot\alpha_i\\
     & & S_i=2\\
  \end{aligned} $}\\
  \bottomrule
\end{tabular}}
\end{table}
Figure \ref{fig:BIAS} shows box-plots of the bias for the
different estimates for the log median ratios. 
Each boxplot shows the realized bias of the posterior estimated log
ratios over 50 repeat simulations.  
For comparison, we also report estimates under a classical meta-regression.
The classical meta-regression results show high variability in the
estimates across simulations, which is substantially reduced by the
proposed model-based inference.  This is due to shrinkage towards
the prior and the borrowing of strength across similar cohorts.
The increased precision is achieved despite the greater flexibility and
in the absence of parametric assumptions in the multivariate
$\mvPT$ model.
\RWnewtext{  Figure \ref{fig:BIAS} suggests that this is due to a trade-off with increased average bias over simulations under the proposed approach. 
More summaries of absolute bias across simulations are available
in the on-line supplementary materials.
We avoid comparing results between tumor-agent pairs, as
  the interpretation of such comparisons hinge 
  on the specific simulation truth and sample sizes, and thus
  would require substantially more simulations scenarios.} 
Finally, keeping in mind that the main inference target
is the evaluation of the hypothesis
$M_i^+ > M_i^-$,
we also assessed methods in terms of inference on this hypothesis by
comparing Bayes factors and Bayes factor bounds
\citep{sellke2001calibration} for classical inference.  The latter
are evaluated using p-values from a meta-regression
These results are shown in the on-line supplementary materials.

\section{Results for the cancer immunotherapy
  meta-analysis}\label{sec:Application} 
\begin{figure}
\begin{minipage}{0.4\textwidth}
    \includegraphics[width=\textwidth]{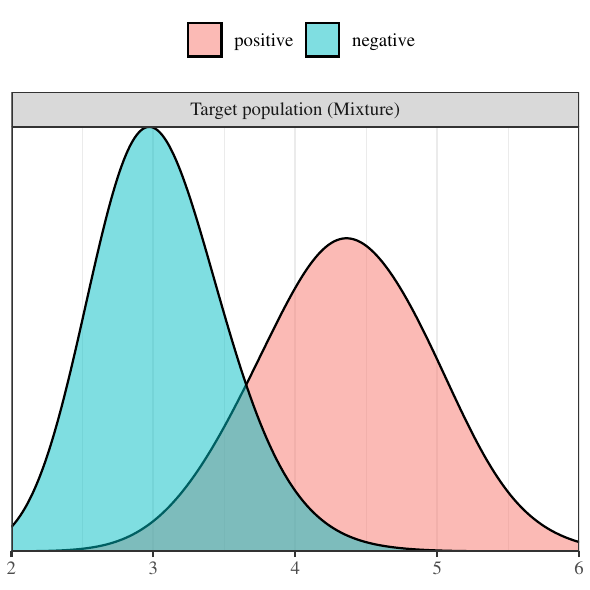}
\end{minipage}\\
\begin{minipage}{0.6\textwidth}
    \end{minipage}
    \begin{minipage}{\textwidth}
\includegraphics[width =\textwidth]{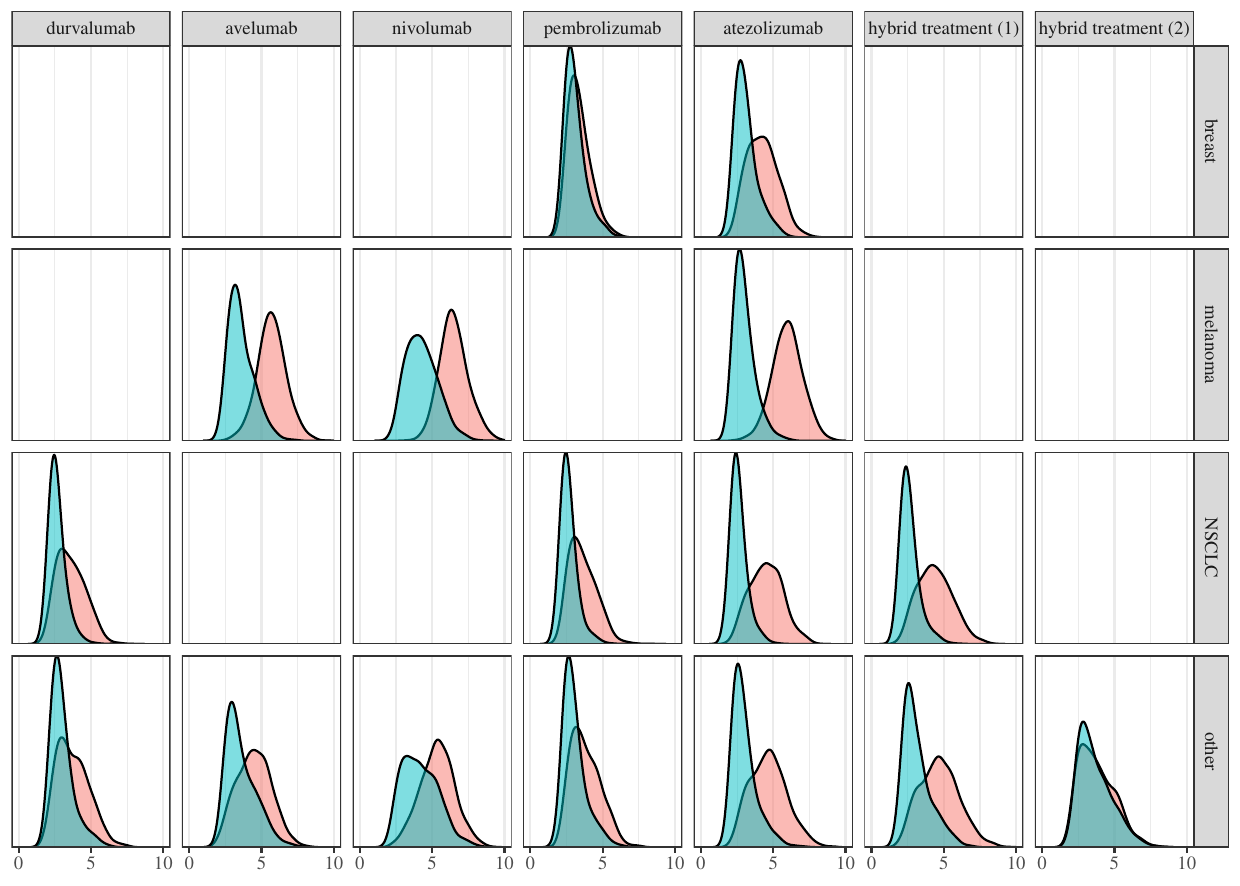}
    \end{minipage}
  \caption{Immunotherapies.
The top panel shows posterior distributions for $M_P^+$ and $M_P^-$
for a hypothetical future study, averaging over study characteristics.   
The lower (small) panels show the same for $M_i^+$ and $M_i^-$ for 
hypothetical future studies, arranged by agent and tumor type.
Here \textit{hybrid treatment} refers to studies with multiple agents,
with (1) for ipilimumab or nivolumab and (2) for pembrolizumab or
nivolumab.
Empty facets indicate there was no study for the tumor agent pair.}
\label{fig:future_med}
\end{figure}

Overall the results support a recommendation for including relevant
biomarkers in the design of cancer immunotherapy studies. For almost
all tumor-agent pairs we find high posterior probability for higher
median PFS
for marker-positive than for marker-negative patients, i.e, for
 $M_P^+>M_P^-$.
Here $P=\sum \pi_i\ G_i$, for a mixture over all future studies with
maker-positive and -negative status, respectively, as described in
Section \ref{subsec:inf}.
Figure \ref{fig:future_med} reports inference for hypothetical future studies,
arranged by combinations of tumor and agent (small panels)
and overall (large panel). Omitted minor
covariates (except for biomarker status) are fixed at the most common
observed level. \textcolor{black}{ Web Figure 6} in the supporting
materials reports the same as Figure \ref{fig:future_med}, but for
$\Mbar_P^+$ and $\Mbar_P^-$.

Our meta-analysis of PFS in immunotherapy studies supports
the hypothesis that biomarkers can be useful to identify
patients who will benefit from
 immune checkpoint inhibitors.  
However, both, the effect size and confidence in the results vary
depending $\xb_i$.
In particular, there is evidence for an interaction of agent and tumor
type.
The results are strongest for melanoma and robust for non-small cell lung
cancer (NSCLC).
For breast cancer (only two studies) and cancers classified as
``other'' results are concordant with clinical practice but with less
strong evidence. 
We find a protective effect for most agents.  The
results for atezolizumab and avelumab are the most consistent across
tumors.  In contrast, for ipilimumab or nivolumab (\textit{hybrid
treatment}) and pembrolizumab, the data does not provide equally
strong evidence.  Posterior credible intervals (one-sided) are
 available in the online supplementary materials. 
Results on pembrolizumab and breast cancer conflict with
clinical experience that would suggest an effect. 
This is likely due to the sample sizes of the
cohorts in the study reported by Adams et. al. (2019).  
These cohorts report equally low estimates
for both medians, $M_i^+$ and $M_i^-$, with large sample sizes
($105$ marker-positive and
$67$ marker-negative patients), which implies higher shrinkage of the
medians of future cohorts with breast cancer and
pembrolizumab towards these point estimates.  This observation
highlights the importance of the covariance function and its local
behavior.

\begin{figure}
    \includegraphics[width = \textwidth]{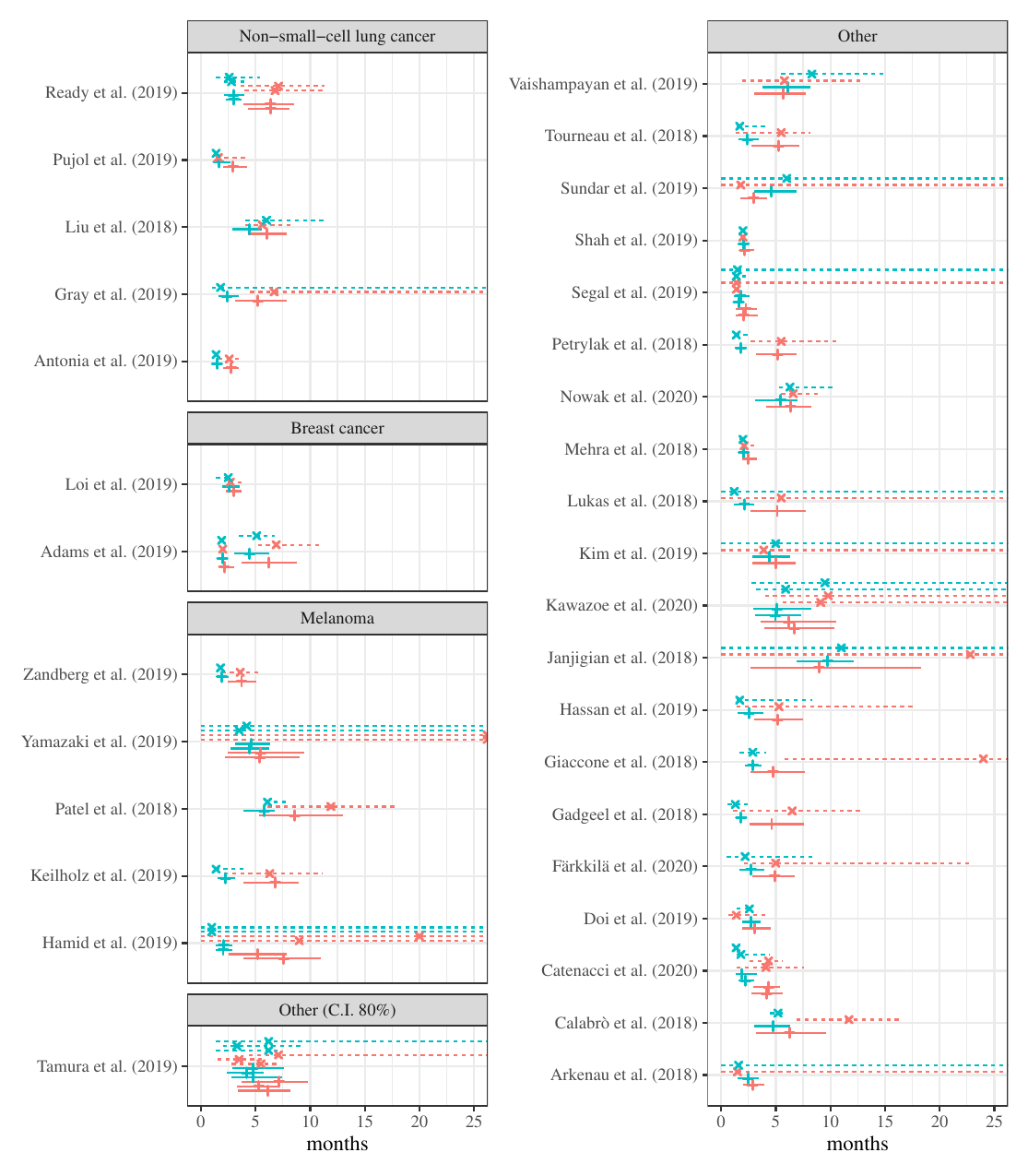}
    \caption{Immunotherapies: Summary of posterior inference for
      the included studies. Intervals represented with $+$ and solid lines
      are credible intervals for median PFS. Intervals
      represented with $\times$ and dotted lines are
      confidence intervals reported in the original articles.
      \RWnewtext{Only for original summaries,} non-observed lower limits are marked with
      0 and non-observed upper limits with $+\infty$. 
      The confidence level for all intervals is 95\%
      unless otherwise specified. 
      \RWnewtext{Marker-positive cohorts are in red and marker-negative cohorts are in blue.}
      \label{fig:obs_med}} 
\end{figure}

Figure \ref{fig:obs_med} summarizes inference for
the original studies included in \cite{fountzilas2023correlation}.  \ech
For each study, the figure compares  
posterior credible intervals for $M_i^+$ and $M_i^-$
together with the confidence intervals in the original papers.
For most studies, the posterior credible intervals for 
$M_i^+$  are higher than those for $M_i^-$.
In particular, \RWnewtext{ for some studies that reported $M^+_i <
M^-_i$ in the original papers, posterior inference switched the order of the point estimates borrowing strength across all cohorts }
(Liu et al. 2018, Tamura et al. 2019, Segal et al. 2019, Kim et
al. 2019, Doi et al. 2019, and 
Arkenau et al. 2018 in Fig. \ref{fig:obs_med}).
The reverse only occurs in one case, for Janjigian et
al. (2018), which is a study reporting extreme values for the medians,
under unusually high and unequal sample sizes (100 marker-positive and
289 marker-negative patients, respectively).

\section{Conclusions}
We developed a non-parametric Bayesian approach for meta-analysis with
event time outcomes.
Inference combines information from all studies in the meta-analysis.
The approach uses weakly informative priors based on clinical expert
judgment regarding the relationship between
 different studies. 
 A simulation study shows that for realistic sample sizes
and data structure inference under the proposed approach compares
favorably with standard methods.  This is especially true for tumor
types and agents that are less commonly observed in the original
studies. This is achieved by borrowing strength across all studies.

One limitation of the proposed non-parametric approach is that it is
restricted to event-time outcomes.
A possible future development would be the inclusion of multiple
endpoints.  Many of the studies in our data include in addition to PFS
also summaries for objective response rate (OR), and overall survival (OS).  Sharing
information on multiple event time endpoints, like PFS and OS, is
easily accommodated by treating them as separate, but highly
correlated, cohorts. The inclusion of binary endpoints like OR in a
joint model would require model extensions to include a parametric
submodel.

\section*{Supplementary Materials}
Pseudo code, Web Tables and Figures referenced in Sections \ref{sec:mvPT1}, \ref{sec:sim} and \ref{sec:Application} and data and code to reproduce the results of the Sections  \ref{sec:sim} and \ref{sec:Application} are available with this paper at the Biometrics website on Oxford Academic.

\section*{Acknowledgement}
First author was supported in part by the ``Dipartimenti Eccellenti 2023-2027'' ministerial funds (Italy). 
Last author was supported in part by National Science Foundation grants DMS 1952679.

\section*{Data Availability}
The data, code and the simulation results that support the findings in this paper are also available on GitHub at link \url{https://github.com/GiovanniPoli/mvPTgp}.

\bibliographystyle{apalike}
\bibliography{references}

\appendix

\section{Posterior update via Pòlya Gamma model augmentation.}

Let $N_{\veps_d}\ii$ denote the number of observations in the interval
$B\ii_{\veps_d}$.
The posterior distribution for ${\Zb_{\veps_d0}} =
\{Z\ii_{\veps_d0}\}_{i=1}^I$ depends on $N\ii_{\veps_d}$ and
$N\ii_{\veps_d 0}$.  The conditional posterior distribution closely
resembles a logistic regression.  Let $A_{\veps_d}$ denote the set of
all cohorts $i$ for which at least one observation is recorded in
$B\ii_{\veps_d}$, and split ${\Zb_{\veps_d0}}$ according to
$A_{\veps_d}$ into $\Zb_{\veps_d0} = {\Zb_{\veps_d0}^1}\cup
{\Zb_{\veps_d0}^\varnothing}$.
Noting that only studies $i \in A$ contribute likelihood factors, 
this implies
\begin{equation*}
\begin{split}
p({\Zb_{\veps_d 0}\mid\cdot})
\propto & \left[\prod_{i\in A_{\veps_d}} \frac{\Big(\exp\Big\{Z_{\veps_d0}\ii\Big\}\Big)^{N\ii_{\veps_d0}}}{\Big(1+\exp\big\{Z_{\veps_d0}\ii\big\}\Big)^{N\ii_{\veps_d}}} \right]
p({\Zb_{\veps_d0}^1})\ p(\Zb_{\veps_d0}^\varnothing\mid \Zb_{\veps_d0}^1).
\end{split}
\end{equation*}
The data augmentation strategy of \cite{polson2013bayesian} is then
implemented as follows:

\begin{itemize}
\item[(i)] Sample ${\omega_i}$ from $p(\omega_i\mid  \bm{Z^1_{\veps_d 0}})$ for each $i\in A_{\veps_d}$ independently using a Pòlya gamma.
\item[(ii)] Update $Z_{\veps_d0}\ii$
for each $i\in A_{\veps_d}$ sampling from  $p(\Zb_{\veps_d0}^1 \mid \bm{\omega})$.
    \item[(iii)] Update $Z_{\veps_d0}\ii$ for each $i\not\in A_{\veps_d}$ by sampling from
      $p(\Zb_{\veps_d0}^\varnothing\mid \Zb_{\veps_d0}^1)$. The latter takes the form of a
      conditional multivariate normal.  
\end{itemize}

\section{Posterior update of (latent) counts}

Consider a generic cohort $i$ with (known) sample size $N$.
Omitting the study-specific indices $i$ and ${(i)}$,  let
$\Cb_{1:N}$ and $\Tb_{1:N}$ denote the unknown patient level censoring and event times.
We first consider the conditional distribution of the complete data, assuming that
both censoring and event time distribution are known. Let $H(C_j)$
denote the earlier, and $G(T_j)$ the latter, and let $\sbf$ denote
the triple $(\ell,m,h)$ implied by $\Tb_{1:N}, \Cb_{1:N}$ and $\sbf^o$
the observed data. Then conditioning on all currently imputed
parameters, including in particular $G$ itself, and the observed data
we have 
\begin{equation*}
\begin{split}
  p\left(\Tb_{1:N}, \Cb_{1:N} \mid \cdot\ \right)
\propto\ & p\left(\sbf^o \mid\Tb_{1:N}, \Cb_{1:N}\right)\prod_{j=1}^N  G\left(T_j\right)H(C_{j})
= \II(\sbf=\sbf^o)\prod_{j=1}^N  G(T_j)\, H(C_{j}).
\end{split}
\end{equation*}
We construct Markov chain Monte Carlos simulation to  impute 
$(\bm{C_{1:N}, T_{1:N}})$ using two transition probabilities defined
with the following proposal distributions. Proposal $Q_1$  proposes a
new event and censoring time for  two randomly selected 
subjects $r$ and $s$ in the cohort.
 We use $T$ and $C$ without superscript for the proposal, and
$T^{o}, C^{o}$ for the currently imputed values. 
$ 
Q_1(\Tb_{1:N}, \Cb_{1:N}\mid \Tb_{1:N}^{o}, {\Cb_{1:N}^{o}})= H\big(C_s\big)
    G\big(T_s\big)\cdot H\big(C_r\big) G\big(T_r\big) / (n(n-1))
$ 
implying acceptance probability
$$  \alpha = \min\left\{1, \frac{\mathbb{I}(\sbf=\sbf')\prod_{j=1}^N
  G\left(T_j\right)H(C_{j})}{\prod_{j=1}^N
  G\left(T_j^{o}\right)H(C^{o}_{j})}\
    \frac{H\big(C_s^{o}\big)
      G\big(T_s^{o}\big)\times H\big(C_r^{o}\big) G\big(T_r^{o}\big)}
    {H\big(C_s\big)
    G\big(T_s\big)\times H\big(C_r\big) G\big(T_r\big)} \right\}
     =\ \II(\sbf=\sbf^o).
     $$
Proposal $Q_2$ randomly selects a subject s and proposes to
\textit{`flip'} the censoring indicator $\delta_s=\mathbb{I}{(T_s<C_s)}$ and
thereby changing the censoring pattern $\bm{\delta}$.
 Let 
$ \Tt_s=\min \{ T_s, C_s\}$ denote the implied observed times,
and let
$\Tt_{(s-1)}$ and $\Tt_{(s+1)}$ denote the largest observed time
less than $\Tt_s$ and the smallest one greater than $\Tt_s$, respectively
(all quantities are known by conditioning on the currently current
imputed values).
Then
\begin{multline}
Q_2(\Tb_{1:N}, \Cb_{1:N}\mid \Tb_{1:N}^{o}, {\Cb_{1:N}^{o}})
=\frac1N \left[
  H\left(C_s\mid\Tt_{ (s-1) }^{o}< C_s  <\Tt_{(s+1)}^{o}\right)
  G\left(T_s\mid C_{s}<T_s,C_s\right)
\right]^{\delta_s^{o}}\\
\times  \left[
  G\left(T_s\mid\Tt_{ (s-1) }^{o}< T_s <\Tt_{(s+1)}^{o}\right)
  H\left(C_s\mid T_{s}<C_s, T_s\right)
\right]^{1-\delta_s^{o}} \\
=\frac1N \left[
  \frac{H\left(C_s\right)}
       {H\left(\Tt_{ (s-1) }^{o}< C_s <\Tt_{(s+1)}^{o}\right)}
  \frac{ G\left(T_s\right)}
       { G\left(T_s>C_{s}\mid C_s \right)} 
\right]^{\delta_s^{o}}  \\ \times
\left[
    \frac{G\left(T_s\right)}
         {G\left(\Tt_{ (s-1) }^{o}< T_s <\Tt_{(s+1)}^{o}\right)}
    \frac{H\left(C_s\right)}
    {H\left(C_s>T_{s}\mid T_s\right)}
  \right]^{1-\delta_s^{o}}.
  \nonumber
\end{multline}
This implies the  acceptance probability $\alpha_2 = \min(1,r)$ with
$$
r = \II(\sbf=\sbf^o)\
\begin{cases}
  \frac{
 G(\Tt_{ (s-1) }^{o}< T_s <\Tt_{(s+1)}^{o})\, H(C_s>T_{s}\mid T_s^{new})}
{H(\Tt_{ (s-1) }^{o}< C_s <\Tt_{(s+1)}^{o})\, G(T_s>C_{s}\mid C_s^{o})}\
\
  & \mbox{ if } \delta^o_s = 1\\[.25cm]
  \frac{
 H(\Tt_{ (s-1) }^{o}< C_s <\Tt_{(s+1)}^{o})\, G(T_s>C_{s}\mid C^{new}_s)} 
{G(\Tt_{ (s-1) }^{o}< T_s <\Tt_{(s+1)}^{o})\, H(C_s>T_{s}\mid T_s^{o})}
    & \mbox{ if } \delta^o_s = 0.
  \end{cases}
  $$
We divided the cohorts into three groups. (1) Cohorts with known
censoring pattern (i.e., no censoring or censoring only after the last
event time); (2) cohorts with reported number of events or number of
 censored outcomes;  (3) cohorts for which only the sample size is
recorded. 
For (1) counts are not updated since they are known. 
For (2) the filp step is never proposed. 
For (3) one of the two proposals is randomly selected each time. 

To initialize $\Tb_{1:N}$ and $\Cb_{1:N}$, we start with a censuring
pattern where all event times are observed and we followed the
following argument.
Let $T_{(j)}$ denote the order statistic for $\Tb_{1:N}$.
We can then consider a Kaplan-Meier plot with the index $j$ on the
horizontal axis (that is, plotting against the indices of the order
statistic instead of the unknown actual times). We plot the
estimated survival function and $(1-\alpha)$
confidence interval bands (still plotted against $j$).
Assuming w.l.o.g. $N=2k+1$, and assuming
that the recorded data $\sbf=(\ell,m,h)$ are determined as
the intersections of the three curves with the 0.5 threshold we can
then identify the data $(\ell,m,h)$ as $(T_{(L)},T_{(k+1)},T_{(H)})$.
The remaining data $T_j$ are generated from $G$, subject to the given
$(T_{(L)},T_{(k+1)},T_{(H)})$. That is, we generate $L-1$ event
times $T_j \sim G \cdot \II(T_j<T_{(L)})$ etc.

\section{Correlation Matrix}
Recall the construction of $R(\xb,\xb')$ in \S\ref{sec:cov}. 
Consider two cohort-specific covariate vectors $\bm{x}$ and $
\bm{x}'$,  and denote with $k_1(\xb,\xb')$ the similarity score
based on matching covariates. 
\begin{itemize}
    \item $k_1(\xb , \xb')$ be a linear kernel with coherent weights
      obtained using the one-hot encoding of the categorical
      variables.
    \item  $k_2(\xb , \xb')$  be an indicator for $\bm{x}$ and $
      \bm{x}'$ having matching marker status. 
    \item  $k_3(\xb , \xb')$  be an indicator for $\bm{x}$ and $ \bm{x}'$ sharing the same study.
    \item  $k_4(\xb , \xb')$  be the identity kernel, i.e., an
      indicator for $\bm{x}$ and $ \bm{x}'$ referring to the same
      cohort. 
\end{itemize}
We combine the kernels to obtain  the covariance functions described
in \S\ref{sec:cov}: 
\[R(\bm{x},\bm{x}')=\frac{k_1(\bm{x},\bm{x}')\cdot k_2(\bm{x},\bm{x}')
    +  k_3(\bm{x},\bm{x}')+ k_4(\bm{x},\bm{x}')}{8}.\]
$\bm{R}$ is positive definite since the sum and the product of positive semi-definitive kernels is positive semi-definite and $k_4(\xb , \xb')$  is positive definite.

For other choices of building the similarity score the construction $R(\bm{x},\bm{x}')$ might not be p.d. In that case, it is always possible to scale $k_4(\xb , \xb')$ by some $c>0$ to ensure R to be p.d
This is the case due to the following result. If $A$ is a symmetric $(I\times I)$ matrix, then there is a $c>0$ such that $B=\left[A+c\cdot I_{I\times I}\right]$ is p.d. The result is 
easy to prove by considering the normalized eigenvalues of B.  See
also \cite{rasmussen2003gaussian} for a discussion of covariance functions,
keeping in mind that the requirement for p.d. covariance matrices is
only needed for the $I$ cohorts under consideration. 

\end{document}